\documentclass[12pt]{iopart}
\usepackage{epsfig,amssymb}
\begin{document}
\title[Declination dependence of the cosmic-ray flux at extreme
energies]{Declination dependence of the cosmic-ray flux at extreme energies}

\author{D.\,S.\,Gorbunov, S.\,V.\,Troitsky
}

\address{
Institute for Nuclear Research of the Russian Academy of
Sciences,\\
60th October Anniversary Prospect 7a, 117312, Moscow, Russia
}
\ead{gorby@ms2.inr.ac.ru, st@ms2.inr.ac.ru}
\begin{abstract} We study the large-scale distribution of the arrival
directions of the highest energy cosmic rays observed by various
experiments. Despite clearly insufficient statistics, we find a deficit
of cosmic rays at energies higher than $10^{20}$~eV from a
large part of the sky around the celestial North Pole.  We speculate on
possible explanations of this feature.
\end{abstract}

\pacno{98.70.Sa}

\maketitle

\section{Introduction}
The physics of the highest-energy cosmic rays continues to
attract significant interest of both particle physicists and
astrophysicists.  Experimental data allow to determine energies
and arrival directions of the cosmic particles and may give some
hints about their nature.  At energies of order $10^{19}$~eV, spectra
measured by different experiments are in a good agreement modulo
overall normalization. Contrary, two major experiments disagree about
the observed flux and shape of the spectrum at highest energies,
$E\gtrsim 10^{20}$~eV: the data from the HiRes experiment exhibit the
suppression of the flux, the so-called GZK feature~\cite{G,ZK}, while
the AGASA experiment does not see the suppression.
Arrival directions of
the cosmic rays with energies $E\gtrsim10^{19}$~eV are distributed
isotropically over the sky (see, for instance, Ref.~\cite{AGASA1}) and
clustered at small angles \cite{AGASA1,clust1,clust2}.  In this note,
we focus on the highest energy region, $E\gtrsim10^{20}$~eV, and
demonstrate that the current data exhibit a trend to non-uniform
distribution of the arrival directions.  Namely, there is a deficit of
events from a large region around the celestial North pole.

For the most energetic cosmic rays observed by AGASA, the
absence of particles coming from the North has been pointed out
in Refs.~\cite{AGASA1,AGASAphotons,Bedn,Stanev}.
However, the distribution of the
arrival directions of these AGASA events is consistent with
isotropy \cite{AGASAwww}, given the non-uniform exposure. To test
the conjecture of anisotropy (and also to improve statistics,
still very poor), we include the results from other Northern
hemisphere experiments in our analysis.
We do observe a non-uniform distribution of events with
probability of this anisotropy to occur as a result of a chance
fluctuation about one percent; this would correspond to two
standard deviations for the Gaussian distribution. If confirmed by better
statistics, this observation might mean that at highest energies, a new
component appears in the cosmic ray flux. Indeed, it is widely believed
now that at $4\cdot 10^{19}~{\rm eV} \lesssim E\lesssim 10^{20}~{\rm eV}$,
the dominant part of the cosmic rays are protons \cite{TTsub,Ber-p} from
active galactic nuclei (in particular, BL Lac type objects suggested
recently as the source candidates \cite{BL1,Uryson,BL2}).  Due to the GZK
effect, this component cannot explain even the most conservative HiRes
flux at  $E\gtrsim 10^{20}~{\rm eV}$ (see Refs.~\cite{Semikoz,Kalashev}
for a quantitative analysis). We will see that indeed, the anisotropy
becomes significant at energies $E\gtrsim 10^{20}~{\rm eV}$.

In the rest of the paper, we discuss in detail the datasets
used,
analyze the declination
distribution of the
highest-energy events detected by
Northern-hemisphere experiments,
present the results of the Monte Carlo simulations
to estimate the chance probability of the observed anisotropy and
study at  what energies the anisotropic component
becomes significant. We discuss possible ways to explain the effect
and demonstrate its irrelevance to the AGASA/HiRes
discrepancy. We emphasize that the number of events is too small
to make a definite conclusion about the anisotropy and briefly
discuss prospects for larger statistics.

\section{The cosmic-ray sample}
Our basic data set (hereafter, Set I) consists of the cosmic rays with energies
higher than $10^{20}~{\rm eV}$ observed by all experiments in the
Northern hemisphere.
The choice of the $10^{20}$~eV cut is determined by the
availability of data from
fluorescent detectors at $E>10^{20}~{\rm eV}$ only.
However, it is roughly consistent with the expected energy at which the
super-GZK component could start to dominate. Alternatively, we consider a
set of cosmic rays with $E> 4\cdot 10^{19}~{\rm eV}$ observed by
Volcano Ranch, Yakutsk and AGASA (Set II; arrival directions of events
with $E<10^{20}~{\rm eV}$ are unpublished for other Northern-hemisphere
experiments). The latter set is useful in determination of the "critical"
energy at which the new, anisotropically distributed, component becomes
important. As it will be demonstrated below, this analysis points to
$E\approx 10^{20}~{\rm eV}$ as the critical energy.

When comparing data from different experiments, one should be careful
about energy normalization. It is widely believed that at $E \sim
10^{19}~{\rm eV}$, where the shapes of the spectra measured by
different experiments agree quite well, the difference in overall
normalization is due to systematic errors in the energy determination. One
possibility is to introduce correcting
factors for energy values in such a way that the total flux measured at $E
= 10^{19}~{\rm eV}$ by different experiments would coincide, within one
standard deviation, with the flux measured by a given (no matter which one)
reference detector
(in our
study, we normalize all fluxes to HiRes data because otherwise the HiRes
dataset is not complete).
The study of the dependence on the choice of the
reference detector is completely equivalent to the study of the dependence
on the "critical" energy performed in \sref{4}.
Quantitatively, this rescaling depends crucially on the assumed
spectral index. We checked, however, that the effect discussed in this
paper is insensitive to this choice: different normalizations (for spectral
indices between two and four) do not change the result -- absence of events
with high declinations -- compared to the case of no
rescaling.
For completeness, we report here (see Table 2) both the results obtained
without energy normalization and with renormalizaton assuming spectral
index three. In the latter case, the corrected energy $E'=\eta E$, where
$E$ is the reported energy of an event and $\eta=(J_{\rm ref}/J)^{1/2}$
(see, for instance, Ref.~\cite{Stanev}). Here, $J_{\rm ref}$ and 
$J\equiv dN/dE$ are the cosmic ray fluxes 
measured by the reference detector and the detector 
under consideration, respectively. The fluxes $J$ at $E \sim 10^{19}~{\rm
eV}$ and normalization factors $\eta$ are listed in Table~1 together with
the information required to calculate and compare exposures of the
experiments and with references.
\begin{table}
\caption{\label{table:exp}Cosmic ray experiments. (1): name; (2): flux measured
at
$10^{19}$~eV (in units of $10^{-33}$~m$^{-2}$ s$^{-1}$ sr$^{-1}$
eV$^{-1}$) and reference; (3): energy rescaling factor; (4): geographic
latitude (in degrees);
(5):
total exposure (in units of $10^{16}$~m$^2$ s sr) (for the
cosmic rays with energies above $10^{19}$~eV; for AGASA
different exposures for published data above $10^{19}$~eV and
$10^{20}$~eV; for Haverah Park, Fly's Eye and HiRes -- at $10^{20}$~eV)
and reference.
}
\lineup
\begin{indented}
\item[]\begin{tabular}{@{}ccccc}
\br
Experiment&
$J$&
$\eta$&
$B$ &
$A$
\\
(1)&
(2)&
(3)&
(4)&
(5)
\\
\mr
Volcano Ranch &$3\pm 1$ \cite{VR}&0.95&32& 0.2 \cite{NW}\\
Haverah Park  &2.2 \cite{NW}&0.90&54&0.9 \cite{NW}\\
Yakutsk       &$4.3\pm 0.6$ \cite{YakICRC03}&0.70&62&1.8
\cite{YakICRC03}\\
Fly's Eye     &$2.5\pm 0.3$ \cite{Bird-spectrum}&0.89&40&2.6
\cite{Bird-spectrum}\\
AGASA, $>\!\!10^{20}$\,eV&      &0.85&36&5.3 \cite{AGASA-ICRC03}\\
AGASA, $<\!\!10^{20}$\,eV&$2.7\pm 0.2$ \cite{AGASA-ICRC03}&0.85&36&4.0
\0\cite{AGASA1}\\
HiRes I mono &$1.6\pm 0.2$ \cite{H1}&1.00&40&6.5 \cite{H1}\\
HiRes II mono&$1.6\pm 0.3$ \cite{H1}&1.00&40&0.7  \cite{H2}\\
HiRes stereo &$2.2\pm 0.2$ \cite{stereo-ICRC}&0.95&40&4.6
\cite{stereo-ICRC}\\
\br
\end{tabular}
\end{indented}
\end{table}

For the data sample, we took the most recent publicly available
data and impose zenith angle cuts of 45$^\circ$ for ground array
experiments and 60$^\circ$ for fluorescent experiments.

Three experiments (Haverah Park,  Yakutsk and HiRes II
in the monocular mode) have considerable exposure at the ultra-high
energies but contributed no events to the Set I (though their exposure
was taken into account).

Energies of the Haverah Park events published
in the Catalog~\cite{CRcat} were reconsidered twice, in
Refs.~\cite{HP-1} and~\cite{HP-2}. According to the most recent
publication~\cite{HP-2}, the energy of the highest event is $E\approx
8.3 \cdot 10^{19}$~eV. Revised event-by-event data were not
published.

The Yakutsk event with the energy $E\approx 1.2\cdot 10^{20}$~eV
has zenith angle $>45^\circ$ and
thus it is not included in Set I
(note that its declination~\cite{Yakutsk} is $\delta\approx
45^\circ$, so its account would only support our conclusions).

Coordinates of the  Volcano Ranch events, both of the only shower
with $E>10^{20}$~eV and of showers with lower energies (Set II) were
taken from Ref.~\cite{CRcat}.

For the Set I, we use the most recent AGASA data from the
experiment's web page~\cite{AGASAwww}. From eleven events,
eight are left after rescaling from $E$ to $E'$. The lower energy data
for the Set II are available for a shorter period of operation,
Ref.~\cite{AGASA1}. This is the reason for smaller exposure used for
the lower-energy data, as indicated in Table~1.

For fluorescent detectors, the data are published for the highest
energy events only. The single Fly's Eye event contributing to
Set I is described in detail in Ref.~\cite{FE}. For the  HiRes
detector in the monocular mode, we take the working period reported in
Ref.~\cite{H1}. The arrival direction of the single event
with $E>10^{20}$~eV registered during that period and passed all
cuts is taken from Ref.~\cite{H-diss}. For the HiRes stereo experiment, we
use the data reported in
Refs.~\cite{stereo-ICRC,Hires2003}\footnote{According to
Ref.~\cite{H-diss}, about 80\% of the stereo events are not included in
the HiRes I monocular data set because of different trigger requirements
and quality cuts, so we consider the exposure of the stereoscopic
observations as one of an independent experiment.}.

\section{Declination-dependent exposure}
Different parts of the sky are seen by various experiments with
different exposures.
In this paper, we use two different approaches: firstly, we model the
dependence of the exposure on declination theoretically; secondly, we use
the actual distribution of the lower-energy events to compare with one of
the highest-energy events.

For a theoretical model of exposure of a ground array, we use the
geometrical differential exposure.  A small plaquette of area $d\sigma$
[sr] on the celestial sphere at the zenith angle $\theta$ is effectively
seen by the area $dA=A_0\cos\theta\,d\sigma$ of a ground array, where
$A_0$ is the maximal aperture. The total exposure for a plaquette
$d\sigma$ can be found by integration over the working time of the
detector. The explicit formulae which result from this integration for a
continuously operating ground array are given in Ref.~\cite{Sommers}. The
exposure does not depend on the right ascension $\alpha$ in this case.  At
energies $E\sim 10^{19}~{\rm eV}$ these
expressions are in rather good agreement with the observed data (see
Fig.~\ref{fig:all}(b)).  To obtain
the normalization factor $A_0$, important for any analysis which involves
data of different experiments, one has to integrate $dA/d\sigma$ over
$d\sigma$ and to compare the resulting total exposure $A$ with the
published value listed in Table~1.

For the fluorescent detectors, this simple geometrical estimate does
not work. A monocular fluorescent detector accepts, at each particular
moment, the cosmic rays uniformly in azimuth and in zenith angle up to
about $50^\circ$, with a relatively sharp drop at larger zenith angles
\cite{H-exp}. In stereoscopic mode, acceptance depends also on the
azimuthal angle. In all cases, fluorescent detectors work on clear
moonless nights only. The information about typical weather and dark sky
availability may be encoded in the dependence of acceptance on sidereal
time~\cite{H-exp}. For the HiRes experiment, we use the zenith angle
dependence of acceptance from Refs.~\cite{H-exp} (monocular detector with
parameters of HiRes) and~\cite{stereo-ICRC} (stereo HiRes detector), the
azimuth angle dependence for the stereo mode from Ref.~\cite{stereo-ICRC}
and the sidereal time dependence from Ref.~\cite{H-exp}.
Though one could expect different zenith angle
dependences of the exposure for different energies, the one we use agrees
quite well with the actual distribution of the HiRes I high-energy
events~\cite{Hires-new}. We are not aware about any published estimate of the
coordinate-dependent exposure for the Fly's Eye experiment in the
monocular mode and (loosely) use the HiRes exposure for it. We present
the results both with and without account of the Fly's Eye event in
Table~2.

\section{Declination distribution of the cosmic rays}
\label{4}
\subsection{Illustration}
\begin{figure}[htb]
\centerline{
\epsfig{file=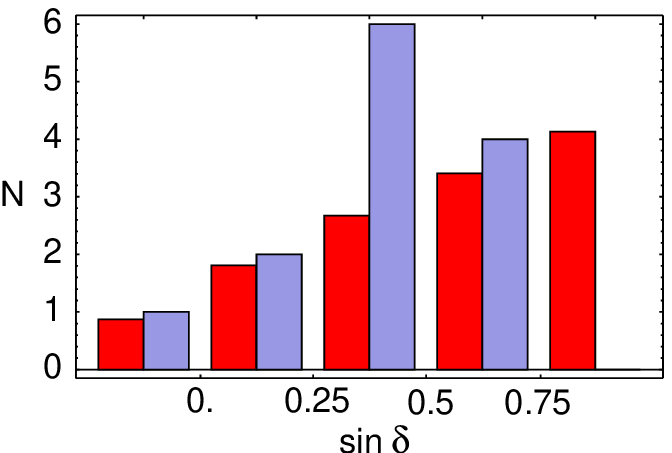, width=0.46\textwidth}
\epsfig{file=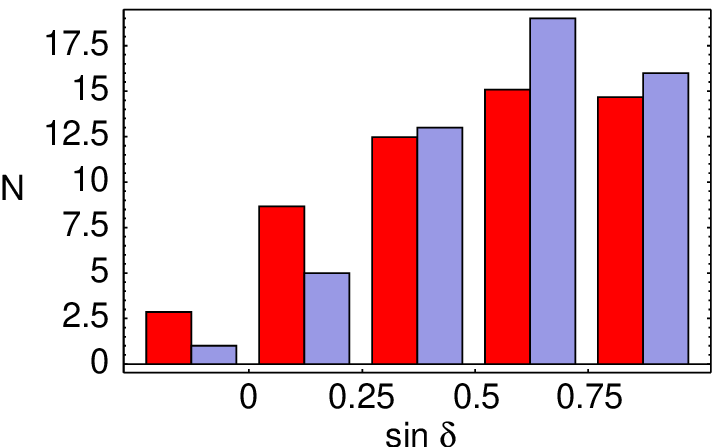, width=0.5\textwidth}}
\centerline{
~~~~(a)~~~~~~~~~~~~~~~~~~~~~~~~~~~~~~~~~~~~~~~~~~~~~~~~~~~~~~~~(b)}
\caption{Expected and observed distributions of
declinations of the cosmic rays: (a) from the Set I ($E'>10^{20}$~eV); (b)
from the Set II ($4\cdot 10^{19}$~eV$<E'<10^{20}$~eV).
\label{fig:all}}
\end{figure}

We are ready to analyze the global distribution of the arrival
directions. To illustrate the anisotropy, we
divided the observed part of the sky into
five bands in declination with equal areas, and
integrated the exposure over these bands.
These exposures were compared then
to the number of observed events, band per band.
The results
for the Set I of cosmic rays are shown in
Figure~\ref{fig:all}(a).
The left (red) bar in each pair corresponds to the exposure per band,
normalized to the total number of events in the sample. It represents
the number of events expected from isotropy. The right
(blue) bar corresponds to the actual number of observed events in the
sample. The deficit of events in the Northern bin is clearly seen.
To estimate this effect quantitatively, we
perform Monte Carlo simulations of
the arrival directions of cosmic rays.

\subsection{Monte-Carlo simulations}
Let us determine the
exposure function
$$
a(\delta)={\cal N} \sum\limits_i \frac{d A_i}{d\delta},
$$
where the sum is taken over all relevant experiments, each one's
total exposure reflected in $A_{0i}$, and ${\cal N}$ is the constant
such that $\max\limits_\delta a(\delta)=1$. The code generates
a value of declination $\delta$ in such a way that the
resulting cosmic rays cover the sky uniformly; then it either
accepts (with the probability $a(\delta)$) or rejects this event
and proceeds to the next one until the total number of accepted
events reaches the number of events in the real dataset~\footnote{For
calculations without energy rescaling, one cannot sum the exposures of
different experiments. Instead, we generate the actual number of observed
events for each experiment in turn; in this way, we do not account the
experiments which observed no events.}. In this way, a sufficient number of
mock sets is produced. The number of sets with no events in the Northern
bin, divided by the total number of sets, determines the probability to
observe the anisotropy by chance. The results are presented in
Table~2.

\begin{figure}
\centerline{\epsfig{file=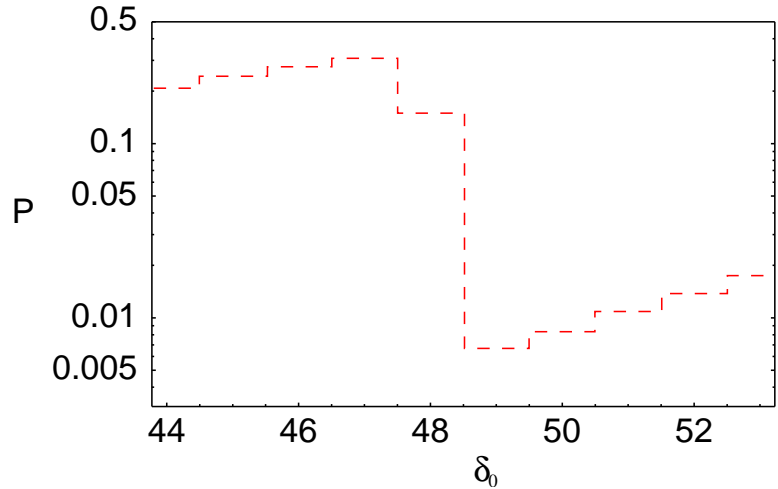, width=0.65\textwidth}}
\caption{Probability {\sf P} to observe the actual number
of events in the Northern bin ($\delta>\delta_0$) as a result of
a statistical fluctuation, versus $\delta_0$ (Set I, $E'>10^{20}$~eV).
\label{fig:bin}}
\end{figure}

To study the stability of our results, we calculated the probabilities
to observe the actual number of events in the Northern bin,
$\delta>\delta_0$,  for
different values of $\delta_0$
(see Figure~\ref{fig:bin}).
The
position of the broad minimum determines the size of the Northern
``zone of avoidance''.

\begin{figure}
\centerline{\epsfig{file=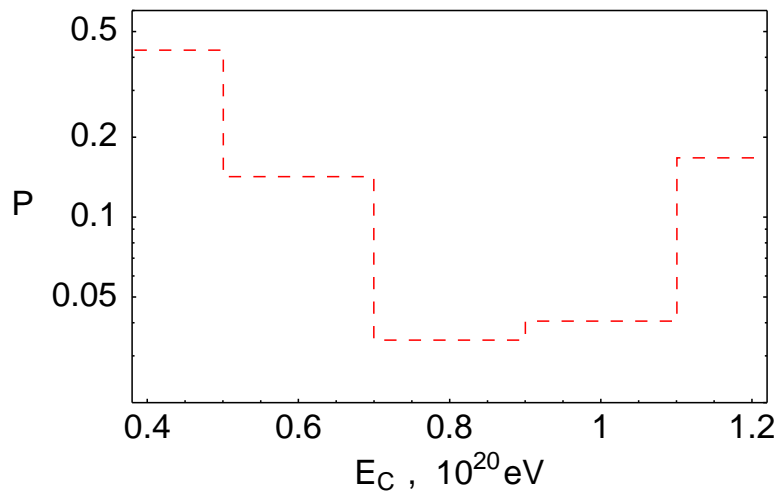, width=0.65\textwidth}}
\caption{Probability {\sf P} to observe the actual
number of events in the Northern bin
as a result of a statistical fluctuation versus the critical energy
$E_c$: data from Set II with $E'>E_c$ are included in the sets.
\label{fig:energy}}
\end{figure}

To understand, at which energy the anisotropically
distributed component becomes important, we use the Set II.
In Figure~\ref{fig:energy},
based on subsets of the Set
II,
we present the probability of the actual number of events to fall in
the Northern bin as a result of a
random fluctuation of the isotropic distribution
for different
cuts on the lower energy
of the cosmic rays in the set. The result
is consistent with our expectations: at lower energies, we confirm the
well-known statement about isotropic distribution; at higher energies,
$E>10^{20}$~eV, protons (from active galactic nuclei) are no longer
dominant in the cosmic ray flux because of the GZK effect, and the new
component is responsible for the observed events, distributed
anisotropically.

Several comments are in order. First, the chance probabilities at
$E'>E_c=10^{20}$~eV presented in Figure~\ref{fig:bin} (Northern bin
corresponds to $\delta_0\approx48.6^\circ$) and in
Figure~\ref{fig:energy} are of the same order, but do not coincide
exactly. The reason is that the sets of experimental data
(Set I and Set II) used in the analysis are different.

Second, because of experimental uncertainties in the determination of
the energies and arrival directions of the cosmic rays, it is not
well-grounded to consider lowest values of probability {\sf P} plotted in
Figures~\ref{fig:bin} and
\ref{fig:energy} as exact numbers: some averaging over the uncertainty
intervals in $\delta$ and $E_c$ should be performed to get more
reliable numbers.  The resulting probabilities will be higher, but
still the isotropy is excluded at the level of about
2$\sigma$ for Gaussian distribution.

Third, the lowest probability in Figure~\ref{fig:energy} corresponds to
the critical energy $E_c\simeq0.8\cdot10^{20}$~eV. With the available data
it is impossible to conclude whether this is the exact energy scale where
the anisotropic component of ultra--high-energy cosmic rays becomes
dominant. Indeed, apart from experimental uncertainties in energy
determination, a systematic error is caused by an arbitrary choice of the
reference detector necessary for a joint analysis of the data from
different experiments (for instance, normalization to one of the ground
arrays instead of HiRes would shift the minimum in Fig.~\ref{fig:energy}
to $\sim10^{20}$~eV). Our choice of the $10^{20}$~eV cut in Set I and of
HiRes as the reference detector was {\it a priory} determined by
availability of data and hence was not adjusted to obtain better results
(in fact, Fig.~\ref{fig:energy} suggests that the best results could be
achieved with another choice). We thus do not introduce statistical
penalty associated with this cut but consider our quantitative results
only as an estimate.

\begin{table}
\caption{\label{table:res}
Probabilities to obtain the observed distribution of declinations from the
isotropic distribution calculated by means of Monte-Carlo simulations of
the lack of events in the Northern bin.
Numbers in parentheses are calculated without the Fly's Eye
event (see Sec.3).
} \lineup
\begin{indented}
\item[]\begin{tabular}{@{}ccc}
\br
Energy rescaling & yes& no\\
\br
$E>10^{20}$~eV versus &&\\
theoretical exposure &1.2\%(1.6\%)&0.9\%(1.4\%)\\
\mr
$E>10^{20}$~eV versus &&\\
$10^{19}~{\rm eV}<E<10^{20}$~eV &1.3\%&1.3\%\\
\br
\end{tabular}
\end{indented}
\end{table}

\section{Conclusions}
Clearly, the number of events in our sample is insufficient to make a
definite conclusion about anisotropy of the arrival
directions. However, the current dataset gives a strong hint that the
deficit of events with energies higher than $10^{20}$~eV and coming
from the region $\delta\gtrsim 50^\circ$ is significant. If confirmed
by future experiments, this fact might suggest that a new component
emerges in the cosmic ray flux at extreme energies. The physics which
could result in the observed distribution of declinations will be
discussed elsewhere; we just mention here possible ways of
explanation. One possibility is that the Northern region of the sky
coincides with the direction to some large-scale cosmic structure
which affects propagation of the cosmic rays at very high energies or
causes inhomogeneous distribution of their sources (at $10^{20}$~eV
this effect is not smeared out by the galactic magnetic field).  Another
option is that the primaries of the cosmic rays with $E\gtrsim
10^{20}$~eV interact with the geomagnetic field and produce showers in
different ways at high and low latitudes. This may lead to a relative
systematic error in the determination of energy between the particles
arriving from different directions. One example of this effect
was discussed in Ref.~\cite{StanevVankov}:
if the primary particle is a photon, then
an electromagnetic cascade develops in the geomagnetic field before
the particle reaches the top of the atmosphere. The observed
superposition of several atmospheric showers mimics a single shower of
the same energy as the primary particle but developed higher, so that
its energy may be underestimated by a ground array.  The current
bounds on the chemical composition of UHECRs \cite{photon1,photon2} do
not constrain strongly the fraction of photonic primaries at $E\gtrsim
10^{20}$~eV; to conclude that the primaries are protons on the base of
the shower profiles and muon counting, one needs
much better statistics than available. However, the data give some
indications to the hadronic nature of primary particles. Primary
protons do not produce pre-atmospheric cascades. Still, the
development of proton-induced showers is affected by the
geomagnetic field: for instance, the separation of muons and
anti-muons is important for modelling~\cite{Dedenko1,Ave1} and energy
estimation~\cite{Dedenko} of inclined showers. The effects of the
geomagnetic field would affect also the distribution of the arrival
directions in azimuth, which will be considered elsewhere.

Future experiments with larger statistics will be able to support or
disfavour the conjecture of anisotropy discussed here. The effects of
the geomagnetic field could be studied with
the Southern site of the Pierre Auger observatory. Clearly, to confirm
or reject the option of a ``favourite direction'' occasionally
coinciding with the North, large detectors in
the Northern hemisphere (such as the second Auger site, the
Telescope Array or the EAS-1000
experiment) or full-sky cosmic observatories (EUSO, OWL, TUS)
would be required.

We stress that the use of currently unpublished data from Haverah
Park, Fly's Eye and HiRes at $E<10^{20}$~eV, as well as of the AGASA
events with zenith angles larger than $45^\circ$, would immediately
enlarge the statistics without awaiting for the future experiments.

Finally, we comment on a recent proposal~\cite{Stanev} that the
discrepancy between the AGASA and HiRes fluxes at highest energies
might be explained by different fields of view. Indeed, the HiRes'
{\it differential} exposure peaks in the Northern region while one of AGASA
has a maximum at $\delta\approx 36^\circ$, the AGASA detector's
latitude. The ``zone of avoidance'' in the North affects the results
of the flux measurements which usually assume the isotropic
distribution of arrival directions. We estimate this effect by a rough
assumption that the cosmic rays with $E\gtrsim 10^{20}$~eV are
distributed uniformly at $\delta< 50^\circ$ but are absent at
$\delta\ge 50^\circ$. With this assumption, a flux measured by HiRes
would be about $40\%$ smaller
than one averaged over all sky, while a flux measured by AGASA decreases by
about $20\%$. Clearly,
this effect cannot eliminate the conflict between the two experiments.

\ack
The authors are indebted to S.~Dubovsky, M.~Fairbairn,
O.~Kalashev, V.~Kuzmin, M.~Libanov, V.~Rubakov, D.~Semikoz,
M.~Teshima, P.~Tinyakov and I.~Tkachev for numerous helpful
discussions.  This work was supported in part by RFBR grant
02-02-17398, by the grants of the President of the Russian Federation
NS-2184.2003.2, MK-2788.2003.02 (D.G.), MK-1084.2003.02 (S.T.) and by the
program SCOPES of the Swiss National Science Foundation, project
No.~7SUPJ062239.
The work of
S.T.\ is supported in part by INTAS grant
YSF 2001/2-129 and by a fellowship of the "Dynasty" foundation
(awarded by the Scientific Council of ICFPM).

\end{document}